\documentclass[journal]{IEEEtran}

\ifCLASSINFOpdf
\else
\fi

\usepackage{amssymb}
\usepackage{amsmath}
\usepackage{color}
\usepackage{graphics}
\usepackage{graphicx}
\usepackage[tight,footnotesize]{subfigure}
\usepackage{algorithm}
\usepackage{algorithmicx,algpseudocode}

\usepackage{multirow}
\usepackage{cite}
\usepackage{epstopdf}
\usepackage{textcomp}

\def \R{\mathbb{R}}

\def \D{\mathcal{D}}

\def \U{\mathcal{U}}

\def \Z{\mathbb{Z}}

\newtheorem{theorem}{Theorem}

\newtheorem{lemma}{Lemma}

\begin{document}
%\maketitle

\title{Economic MPC of Nonlinear Systems with Non-Monotonic Lyapunov Functions and Its Application to HVAC Control}

\author{Zheming Wang and Guoqiang Hu
\thanks{The authors are with the School of Electrical and Electronic Engineering, Nanyang Technological University, Singapore 639798 (email: wangzm@ntu.edu.sg, gqhu@ntu.edu.sg).}}

\maketitle

\begin{abstract}
This paper proposes a Lyapunov-based economic MPC scheme for nonlinear sytems with non-monotonic Lyapunov functions. Relaxed Lyapunov-based constraints are used in the MPC formulation to improve the economic performance. These constraints will enforce a Lyapunov decrease after every few steps. Recursive feasibility and asymptotical convergence to the steady state can be achieved using Lyapunov-like stability analysis. The proposed economic MPC can be applied to minimize energy consumption in HVAC control of commercial buildings. The Lyapunov-based constraints in the online MPC problem enable the tracking of the desired set-point temperature. The performance is demonstrated by a virtual building composed of two adjacent zones.
\end{abstract}

%\renewcommand{\abstractname}{Note to Practitioners}
%\begin{abstract}
%Economic MPC is considered as a promising solution for minimizing energy consumption of commercial buildings due to its optimal nature. One of the main advantages is that it is able to handle various hard constraints such as thermal comfort, actuator limitations and safety constraints. However, as economic MPC only considers energy consumption in the objective function, it may not able to track the desired set-point temperature determined by users. To solve this problem, Lyapunov-based constraints are imposed in the proposed economic MPC scheme, although we may lose some economic performance. In practice, the trade-off between economic performance and convergence speed to the desired set-point temperature can be manipulated by the Lyapunov decrease step-size. The proposed economic scheme in this paper is a centralized control strategy. For buildings with multiple zones, we believe it is possible to develop a distributed economic MPC scheme based on the results in this paper. In the future, we will extend the proposed economic MPC scheme to multi-agent systems and consider the application to multi-zone thermal systems for HVAC control.
%\end{abstract}


%%%%%%%%%%%%%%%%%%%%%%%%%%%%%%%%%%%%%%%%%%%%%%%%%%%%%%%%%%
\section{Introduction}
The building sector accounts for almost $40\%$ of  the world’s total end use of energy\cite{ART:L08}. A significant amount of the energy is consumed for comfort control in Heating Ventilation and Air Conditioning (HVAC) systems. In recent years, model predictive control (MPC) have been widely used to minimize energy consumption and costs of HVAC systems\cite{ART:OPJGGSLM12,ART:MBHCBH12,ART:SGMS16,ART:VJZBL16}. The general framework of MPC is to solve an online finite horizon optimization problem having a cost function with constraints on the predicted state and predicted control.

In standard tracking MPC, the cost function is positive definite with respective to some set-point or trajectory and stability can be obtained with appropriate terminal conditions\cite{ART:MRRS00}. However, in HVAC systems, the cost function may be a reflection of the energy consumption or the process economics. This motivates the use of economic MPC in HVAC control as it is able to optimize the process economic performance directly \cite{INP:RAB12,ART:EDC14}. In contrast to tracking MPC, economic MPC uses some general economic cost function which is not necessary positive definite with respective to any set-point or trajectory. However, the drawback is that it is difficult to establish the stability of economic MPC since the stability analysis techniques in tracking MPC are no long valid.

One well-known method to establish the Lyapunov stability of economic MPC is to use a dissipativity condition. The first Lyapunov-like stability analysis is provided in \cite{ART:DAR11} by modifying the economic cost function. A monotonically decreasing Lyapunov function can be constructed on the assumption of strong duality and thus the asymptotic stability of the closed-loop system is obtained. The assumption of strong duality is then generalized by \cite{ART:AAR12} using a dissipativity condition of the system. In these works, only an equality terminal constraint is imposed and there is no terminal cost function. Similar to \cite{ART:DAR11,ART:AAR12}, the works \cite{ART:ARA11,ART:MAAAR14} also use the dissipativity condition to establish the asymptotic stability. The difference is that they use a terminal region constraint instead of an inequality constraint in order to increase the size of the feasible domain and improve the close-loop performance. In \cite{ART:G13}, the terminal constraints can be removed at the cost of using a sufficiently long prediction horizon. However, there is case where the dissipativity condition is not satisfied.

An alternative method for achieving the stability properties of economic MPC is to design Lyapunov-based constraints by using some auxiliary stabilizing controller. Such a method is called Lyapunov-based economic MPC. In \cite{ART:HLC12,ART:EC14}, by the use of Lyapunov-based constraints, the state is enforced within a level set of a given control Lyapunov function and is ultimately bounded in a small region containing the set-point. Similar techniques have been used in \cite{ART:HWS15,ART:Z15,ART:HSY16}, which refer to the Lyapunov-based constraint as the stabilizing or contractive constraint. These constraints will enforce a Lyapunov decrease and steer the state to the exact desired set-point. However, it may be conservative to enforce the Lyapunov function to decrease monotonically. The constraints used in the works mentioned above can be relaxed so that only non-monotonic Lyapunov functions are needed to achieve asymptotic stability. Motivated by this observation, this work aims to design relaxed Lyapunov-based constraints to improve the economic performance of the closed-loop system.

The rest of the paper is organized as follows. This section ends with the notations needed, followed by the next section on the review of preliminary results of standard tracking MPC. Section \ref{sec:main} presents the proposed economic MPC with relaxed Lyapunov-based constraints and non-monotonic Lyapunov functions. The feasibility and stability of the closed-loop system is shown in Section \ref{sec:stability}. The average economic performance of the closed-loop system is discussed in Section \ref{sec:av}. Section \ref{sec:num} discusses the application of the proposed economic MPC to HVAC control. The last section concludes the work.

The notations used in this paper are as follows. Non-negative and positive integer sets are indicated by $\mathbb{Z}^+_0$ and $\mathbb{Z}^+$ respectively with $\mathbb{Z}^M:=\{1,2,\cdots,M\}$ and $\mathbb{Z}_L^M:=\{L,L+1,\cdots,M\}, M \ge L$, $M, L \in \Z^+_0$. Similarly, $\mathbb{R}^+_0$ and $\mathbb{R}^+$ refer respectively to the sets of non-negative and positive real number. $I_n$ is and $n\times n$ identity matrix.  For a square matrix $Q$, $Q \succ (\succeq) 0$ means $Q$ is positive definite (semi-definite). The $p$-norm of $x\in \mathbb{R}^{n}$ is $\|x\|_p$ (the subscript will be omitted for $p=2$) while $\|x\|^2_Q=x^TQx$ for $Q \succ 0$. For a set $S\subset \R^n$ and a point $x\in \R^n$, the distance between the point and the set $S$ is defined as $\|x\|_S:= \inf_{z\in S}\|x-z\|$. A function $\rho:\R^n \rightarrow \mathbb{R}^+_0$ is positive definite respective to $\bar{x}$ if it is continuous, $\rho(\bar{x}) = 0$, and $\rho(x) >0$ for all $x\not=\bar{x}$. Several representations of the states and controls are needed:  $x(t)$, $u(t)$ refer to the state and control of the system at time $t$; $x_k$, $u_k$ are the $k^{th}$ predicted state and control; boldface $\pmb{x}=(x_0, x_1, \cdots, x_{N})$, $\pmb{u}=(u_0, u_1, \cdots, u_{N-1})$ are the collections of the predicted states and predicted controls over the horizon (of length $N$); in situation where the reference to time is needed, $x_k$, $u_k$ can be written as $x_{k|t}$ and $u_{k|t}$. Additional notations are introduced as required in the text.

\section{Preliminaries and problem statement}\label{sec:pro}
Consider the discrete-time nonlinear system of the form
\begin{align}
x(t+1) &= f(x(t),u(t)), ~~ t\in \Z^+_0 \label{eqn:xplus1AB}\\
x(t) &\in X, u(t)\in U,
\end{align}
where $x(t)$ and $u(t)$ are the state and input of the system, and $X\subseteq \R^{n_x}$ and $U\subseteq \R^{n_u}$ are some appropriate state and control constraint sets. The state and input constraints should be satisfied at each time instant. The instantaneous economic performance is measured by a function of the form $l_e:\R^{n_x}\times\R^{n_u}\rightarrow \R$. In addition, this paper also considers an asymptotic state constraint $X_{\infty}\subseteq X$ that only has to be satisfied asymptotically, i.e., $\lim\limits_{t\rightarrow \infty} \|x(t)\|_{X_{\infty}}=0$. In HVAC systems, the set $X_{\infty}$ would be the desired temperature set-points determined by users. Let $(x_s,u_s)$ denote the admissible optimal economic steady state that minimizes the following problem
\begin{subequations}\label{eqn:xsus}
\begin{align}
(x_s,u_s) = \arg &\min\limits_{(x,u)}l_e(x,u) \\
\textrm{s.t.   } \quad & x = f(x,u),x\in X_{\infty}, u\in U.
\end{align}
\end{subequations}
The objective of this paper is to design a control law such that the economic performance is minimized and the closed-loop system will reach the steady state $(x_s,u_s)$ asymptotically. Note that the set $X_{\infty}$ is usually small or even a singleton. This constraint will not be imposed as a state constraint in the MPC problem as otherwise the feasible domain becomes very small. In this paper, we do not assume the dissipativity condition as shown in \cite{ART:AAR12} and the associated Lyapunov-like stability analysis is not used.

This section then reviews some well-known results in standard tracking MPC and other related concepts. The standard MPC problem can be given by
\begin{subequations}
\begin{align}
\mathbb{P}(x):\quad \min\limits_{\pmb{u}} & V(x,\pmb{u})=\sum\limits_{k=0}^{N-1} l(x_k,u_k) + l_f(x_N) \label{eqn:LQcost} \\
\textrm{s.t.   } &x_{k+1} = f(x_k,u_k),x_0 = x,\\
&x_k \in X, u_k \in U,x_N\in X_f,k \in \Z^{N-1}_0,
\end{align}
\end{subequations}
where $N$ is the horizon length, $\pmb{u}:=\{u_0, u_1, \cdots, u_{N-1}\}, \pmb{x}:=\{x_0, x_1, \cdots, x_{N}\}$ are the predicted controls and predicted states respectively, $l(x_k,u_k)$ and $l_f(x_N)$ are the stage and terminal cost functions respectively, and $X_f$ is some appropriate terminal set. For standard tracking MPC, the stage and terminal cost functions are positive definite with respective to $(x_s,u_s)$
\begin{align}\label{eqn:llf}
0 = l(x_s,u_s) \le l(x,u), 0 = l_f(x_s) \le l_f(x), \forall (x,u).
\end{align}
The terminal constraint set is some constraint-admissible invariant set with some stabilizing control law $\kappa_f:X \rightarrow U$ satisfying
\begin{align}\label{eqn:AKx}
f(x,\kappa_f(x)) \in X_f, ~~\kappa_f(x) \in U \textrm{ for all } x \in X_f.
\end{align}
For notational simplicity, let $\U(x):=\{\pmb{u}\in \R^{n_uN}: x_{k+1} = f(x_k,u_k),x_0 = x,x_k \in X, u_k \in U,x_N\in X_f,k \in \Z^{N-1}_0\}$. The feasible domain is defined by $\D:=\{x\in X: \U(x) \not= \emptyset\}$. The following assumptions are needed in the sequel. These assumptions are standard in tracking MPC \cite{ART:MRRS00}.

\textbf{A1.} $(A,B)$ is controllable and $x(t)$ is measurable.

\textbf{A2.} The functions $f(x,u)$ and $l_e(x,u)$ are continuous in $(x,u)$.

\textbf{A3.} The problem (\ref{eqn:xsus}) is feasible and the constraint sets $X\subseteq \R^n$ and $U\subseteq \R^m$ are compact sets that contain $x_s$ and $u_s$ in their interiors respectively.

\textbf{A4.} There exist a compact terminal region $X_f \subseteq X$ and a terminal control law $\kappa_f:X \rightarrow U$ such that (\ref{eqn:AKx}) is satisfied, $\kappa_f(x_s) = u_s$ and $X_f$ contains $x_s$ in its interior.

%The terminal constraint set is some constraint-admissible invariant set satisfying
%\begin{align}
%A_K (x-x_s)+x_s \in X_f(x_s,u_s), ~~K(x-x_s)+u_s \in U \textrm{ for all } x \in X_f(x_s,u_s)\label{eqn:AKx}
%\end{align}
%where $A_K:=A+BK$ is Schur stable and $K$ is some stabilizing control law.

%Usually, $l(x_k,u_k)$ and $l_f(x_N)$ are chosen to be quadratic costs:
%\begin{align}
%l(x_k,u_k)=\|x_k-x_s\|_{Q}^2 + \|u_k-u_s\|_{R}^2, l_f(x_N)=\|x_N-x_s\|_{P}^2
%\end{align}
%for some appropriate matrices $Q,R,P\succ0$ and a scalar $\delta>0$ that satisfy
%\begin{align}
%(A+BK)^TP(A+BK)-P \preceq -(Q+K^TRK) - \delta I_n
%\end{align}
%with some stabilizing $K$. As a result, $V(x,\pmb{u})$ is also a quadratic cost parameterized by $(x,\pmb{u})$.

In economic MPC, the economic cost function $l_e(x_k,u_k)$ is directly employed in the online optimization problem, as shown below.
\begin{subequations}\label{eqn:EMPC}
\begin{align}
\mathbb{P}_{e}(x) :\quad \min\limits_{\pmb{u}} V_e(x,\pmb{u}):=& \sum\limits_{k=0}^{N-1}l_e(x_k,u_k)  \\
\textrm{s.t.   } & \pmb{u} \in \U(x)
\end{align}
\end{subequations}
Suppose the solution of (\ref{eqn:EMPC}) is $\pmb{u}^*:=\{u^*_{0},u^*_{1},\cdots,u^*_{N-1}\}$, the standard MPC control law is given by
\begin{align}\label{eqn:ukappa}
u = \kappa(x) = u^*_{0}.
\end{align}
Since the economic cost is not necessarily a positive definite function with respective to $(x_s,u_s)$, it is not guaranteed that the closed-loop system with the control law (\ref{eqn:ukappa}) will converge to the steady state. One method to ensure the stability of economic MPC is to use Lyapunov-based constraints \cite{ART:HWS15,ART:Z15,ART:HSY16}. The idea follows from the standard MPC arguments on recursive feasibility and stability \cite{ART:MRRS00}. However, it is not necessary to construct a monotonic Lyapunov function to guarantee stability as the induced Lyapunov-based constraints will become restrictive and potentially undermine economic performance. To enhance economic performance, this work considers more relaxed Lyapunov-based constraints and stability can still be guaranteed by non-monotonic Lyapunov functions.

\section{Economic MPC with Lyapunov-based constraints}\label{sec:main}
This section presents an economic MPC scheme where there exists a non-monotonic Lyapunov function which is not decreasing at every time instant. Instead, the Lyapunov function is required to decrease after every few steps to ensure stability. Suppose the control law of the economic MPC is denoted by $u(t) = h(t,x(t))$, the closed-loop system becomes
\begin{align}
x(t+1) = f(x(t),h(t,x(t))).
\end{align}
We aim to design some proper control $h(t,x(t))$ such that there exists a Lyapunov function denoted as $V^*(x(t))$ which satisfies the following property for some given design parameter $m \ge 1$:
\begin{align}\label{eqn:Vtm}
V^*(x(t+m)) \le V^*(x(t)) - \rho(x(t)), t\in \Z^+_0,
\end{align}
where $\rho: X \rightarrow \R^+_0$ is a positive definite function with respect to $x_s$. When $m=1$, it become a monotonic Lyapunov function, which is similar to the works in \cite{ART:HWS15,ART:Z15,ART:HSY16}.
%In the case of $m=1$, the property (\ref{eqn:Vtm}) can be easily obtained using (\ref{eqn:EMPCLy1}). In the following, we consider the case where $m>1$.
\subsection{The modified tracking value function}
Before the economic MPC scheme, a modified tracking value function is needed. Given $(x,\pmb{u})$, define
\begin{align}\label{eqn:Vdelta}
V^{\delta}(x,\pmb{u}) = \sum\limits_{k=0}^{N-1} (l(x_k,u_k)+k\delta(x_k,u_k)) + l_f(x_N)
\end{align}
where $\delta: X \times U \rightarrow \R$ is a positive definite function with respective to $(x_s,u_s)$. The additional terms are used to enforce all the predicted states including the terminal state to approach the steady state $x_s$, as will be shown later. In order to establish the stability results, the following assumption is made concerning the terminal cost $l_f$. This is a modification of the standard assumption on the terminal cost \cite{ART:MRRS00}.

\textbf{A5.} For any $x\in X_f$, the terminal cost $l_f$ satisfies
\begin{align}
l_f(f(x,\kappa_f(x))) - l_f(x) \le &-l(x,\kappa_f(x)) \label{eqn:P}\\
&- (N-1)\delta(x,\kappa_f(x)) - \gamma(x) \nonumber
\end{align}
where $\gamma: X \rightarrow \R^+_0$ is some positive definite function with respective to $x_s$.

%where $Q,R\succ0$ are the design parameters and $P\succ0$ satisfies
%\begin{align}\label{eqn:P}
%(A+BK)^TP(A+BK)-P \preceq -(Q+N\delta I) - K^T(R+(N-1)\delta I)K
%\end{align}
%Since $A+BK$ is Schur stable, the existence of such a $P$ is guaranteed.

Let
\begin{align}\label{eqn:J}
J^{\delta}(x,\pmb{u}) = l(x_0,u_0) + \sum\limits_{k=1}^{N-1}\delta(x_k,u_k) + \gamma(x_N).
\end{align}
The following result is the direct consequence of the definition of the tracking value function $V^{\delta}(x,\pmb{u})$.

\begin{lemma}\label{lem:convergence}
For any $x\in \D$, let $\pmb{u}:=\{u_0, u_1, \cdots, u_{N-1}\}\in \U(x)$ be a feasible control sequence with the associated state sequence $\pmb{x}:=\{x_0, x_1, \cdots, x_{N}\}$. Suppose the successor state is denoted by $x^{+}=f(x_0,u_0)$. Define a control sequence $\pmb{u}^+:=\{u_1,\cdots,u_{N-1},u_{N}\}$ with $u_N =\kappa_f(x_N)$. The following results hold.\\
(i) $\pmb{u}^+\in \U(x^+)$.\\
(ii) $V^{\delta}(x^+,\pmb{u}^+)-V^{\delta}(x,\pmb{u}) \le - J^{\delta}(x,\pmb{u})$.
\end{lemma}
\textbf{Proof of Lemma \ref{lem:convergence}:} (i) This property directly follows from the standard arguments of recursive feasibility of MPC\cite{ART:MRRS00} and hence the proof is omitted.\\
(ii) The state sequence associated with $\pmb{u}^+$ can be given by $\pmb{x}^+:=\{x_1,\cdots,x_N,x_{N+1}\}$ with $x_{N+1}=f(x_N,\kappa_f(x_N))$. Hence, it follows that
\begin{align*}
V^{\delta}(x^+,\pmb{u}^+) = \sum\limits_{k=1}^{N} (l(x_k,u_k)+(k-1)\delta(x_k,u_k)) + l_f(x_{N+1}).
\end{align*}
This implies that
\begin{align}
&V^{\delta}(x^+,\pmb{u}^+)-V^{\delta}(x,\pmb{u}) \nonumber\\
= &l_f(x_{N+1}) + l(x_N,\kappa_f(x_N)) + (N-1)\delta(x_N,\kappa_f(x_N)\nonumber\\
&-l_f(x_N)-l(x_0,u_0) - \sum\limits_{k=1}^{N-1}\delta(x_k,u_k) \nonumber\\
\le & -\gamma(x_N) - l(x_0,u_0) - \sum\limits_{k=1}^{N-1}\delta(x_k,u_k)\nonumber\\
= & -J^{\delta}(x,\pmb{u}) \nonumber
\end{align}
where the inequality follows from (\ref{eqn:P}). $\Box$

\subsection{The monotonic Lyapunov function}
Based on the results in Lemma \ref{lem:convergence}, we can develop an economic MPC scheme with a monotonic Lyapunov function. This is the case where $m=1$ in (\ref{eqn:Vtm}). For some given parameter $\eta_t$, a Lyapunov-based constraint can be imposed to the economic MPC problem (\ref{eqn:EMPC}) at time $t$ as follows
\begin{subequations}\label{eqn:EMPCLy1}
\begin{align}
\mathbb{P}^{1}_{e}(x(t),\eta_t):\quad \min\limits_{\pmb{u}} V_e(x(t),\pmb{u}):=& \sum\limits_{k=0}^{N-1}l_e(x_k,u_k)  \\
\textrm{s.t.   } &\pmb{u} \in \U(x(t)),\\
& V^{\delta}(x(t),\pmb{u}) \le \eta_t.  \label{eqn:Vdeltax0}
\end{align}
\end{subequations}
Suppose the optimal solution of (\ref{eqn:EMPCLy1}) at time $t$ is denoted as $\pmb{u}^*_{t}:=\{u^*_{0|t},u^*_{1|t},\cdots,u^*_{N-1|t}\}$, the control law is
\begin{align}\label{eqn:kappa1eta}
u(t) = \kappa_e^1(x(t),\eta_t) = u^*_{0|t}.
\end{align}
The parameter $\eta_t$ is updated according to the following adaptive law
\begin{align}\label{eqn:etat}
\eta_t = V^{\delta}(x(t-1),\pmb{u}^*_{t-1})- \beta J^{\delta}(x(t-1),\pmb{u}^*_{t-1})
\end{align}
where $\beta\in (0,1]$ is some fixed scalar. The initial value $\eta_0$ is set to be $+\infty$. Hence, at the initial state $x(0)$, $\mathbb{P}^{1}_{e}(x(0),\eta_0)$ will reduce to $\mathbb{P}_{e}(x(0))$. Using $\mathbb{P}^{1}_{e}(x(t),\eta_t)$, the economic MPC scheme with a monotonic Lyapunov function can be described below.
\begin{algorithm}[H]
\caption{The economic MPC with a monotonic Lyapunov function}
\begin{algorithmic}[1]
\State \textit{Initialization}: Set $t=0$, measure the initial state $x(0)$, let $\eta_0 = +\infty$, solve $\mathbb{P}^{1}_{e}(x(0),\eta_0)$ and obtain the optimal solution $\pmb{u}^*_0$ at time $t=0$. Apply the control law (\ref{eqn:kappa1eta}) to system (\ref{eqn:xplus1AB}). Let $t:=t+1$ and go to Step 2.
\State Measure the current state $x(t)$ and determine $\eta_{t}$ according to (\ref{eqn:etat}).
\State Solve $\mathbb{P}^{1}_{e}(x(t),\eta_t)$ and obtain the optimal solution $\pmb{u}^*_t$. Apply the control law (\ref{eqn:kappa1eta}) to system (\ref{eqn:xplus1AB}).
\State Wait for next sampling time, let $t:=t+1$ and go to Step 2.
\end{algorithmic}
\label{algo:EMPC1}
\end{algorithm}

Due to the additional constraints (\ref{eqn:Vdeltax0}), the recursive feasibility and stability of economic MPC can be guaranteed using $V^{\delta}(x(t),\pmb{u}^*_{t})$ as the Lyapunov function. These results are stated in the following theorem.

\begin{theorem}\label{thm:1}
Suppose $x(0)\in \D$ and $\pmb{u}^*_0\in \U(x(0))$ is the optimal solution of $\mathbb{P}^{1}_{e}(x(0),\eta_0)$ with $\eta_0 = +\infty$. For $t\ge 0$, let $\pmb{u}^*_{t}$ denote the solution obtained from $\mathbb{P}^{1}_{e}(x(t),\eta_t)$. Suppose $\eta_t$ is updated according to (\ref{eqn:etat}) and the control law (\ref{eqn:kappa1eta}) is applied to system (\ref{eqn:xplus1AB}). Then, the following results hold.\\
(i) $\mathbb{P}^{1}_{e}(x(t),\eta_t)$ is feasible for all $t\ge 0$.\\
(ii) $\lim\limits_{t\rightarrow \infty}V^{\delta}(x(t),\pmb{u}^*_{t}) = 0$ and  $\lim\limits_{t\rightarrow \infty}J^{\delta}(x(t),\pmb{u}^*_{t}) = 0$. \\
(ii) The closed-loop system (\ref{eqn:xplus1AB}) controlled by the MPC control law (\ref{eqn:kappa1eta}) asymptotically converges to $x_s$.
\end{theorem}
\textbf{Proof of Theorem \ref{thm:1}:} (i) Suppose the predicted state sequence associated with $\pmb{u}^*_0$ is denoted as $\pmb{x}^*_0 = \{x^*_{0|0},x^*_{1|0},\cdots,x^*_{N|0}\}$. The shifted control sequence at time $t=1$ can be given by $\tilde{\pmb{u}}_1 = \{u^*_{1|0},\cdots,u^*_{N-1|0},\kappa_f(x^*_{N|0})\}$. From Lemma \ref{lem:convergence}, we know that $\tilde{\pmb{u}}_1 \in \U(x(1))$ and $V^{\delta}(x(1),\tilde{\pmb{u}}_1) \le V^{\delta}(x(0),\pmb{u}_{0})- J^{\delta}(x(0),\pmb{u}_{0}) \le V^{\delta}(x(0),\pmb{u}_{0})- \beta J^{\delta}(x(0),\pmb{u}_{0}) = \eta_1$. Hence, $\mathbb{P}^{1}_{e}(x(1),\eta_1)$ is feasible. By induction, $\mathbb{P}^{1}_{e}(x(t),\eta_t)$ is feasible for all $t\ge 0$.\\
(ii) The constraint (\ref{eqn:Vdeltax0}) implies that  $V^{\delta}(x(t),\pmb{u}^*_t) \le V^{\delta}(x(t-1),\pmb{u}^*_{t-1})- \beta J^{\delta}(x(t-1),\pmb{u}^*_{t-1})$ for all $t\ge 1$. Hence $\{V^{\delta}(x(t),\pmb{u}^*_t)\}$ is a monotonic non-increasing sequence bounded from below by 0. This implies the existence of $\lim\limits_{t\rightarrow \infty} V^{\delta}(x(t),\pmb{u}^*_t)$ and $\lim\limits_{t\rightarrow \infty} J^{\delta}(x(t),\pmb{u}^*_{t})=0$, which in turn implies that $\lim\limits_{t\rightarrow \infty} V^{\delta}(x(t),\pmb{u}^*_t) = 0$ as $J^{\delta}$ is positive definite with respective to the steady state.\\
(iii) This result has already proved in (ii).

\subsection{The proposed economic MPC scheme}
The formulation above is now extended to an economic MPC scheme with non-monotonic Lyapunov functions. This means that we consider the case where $m>1$ in (\ref{eqn:Vtm}). Given parameters $\xi_t$ and $\zeta_t$, the proposed economic MPC formulation is given by
\begin{subequations}\label{eqn:EMPCLym}
\begin{align}
\mathbb{P}^{m}_{e}(x(t),\xi_t,\zeta_t):\quad &\min\limits_{\pmb{u}} V_e(x(t),\pmb{u}):= \sum\limits_{k=0}^{N-1}l_e(x_k,u_k)  \\
\textrm{s.t.   } &\pmb{u} \in \U(x(t)),\\
& V^{\delta}(x(t),\pmb{u})\le \xi_t, \label{eqn:Vx0uVtm}\\
& V^{\delta}(x(t),\pmb{u})-\beta J^{\delta}(x(t),\pmb{u}) \le  \zeta_t. \label{eqn:V0l0Vtminus12}
\end{align}
\end{subequations}
where $\beta \in (0,1]$. As will be discussed soon, the constraint (\ref{eqn:V0l0Vtminus12}) is needed to ensure the recursive feasibility. Suppose the optimal solution of (\ref{eqn:EMPCLym}) at time $t$ is denoted as $\pmb{u}^*_{t}:=\{u^*_{0|t},u^*_{1|t},\cdots,u^*_{N-1|t}\}$, the control law is
\begin{align}\label{eqn:kappam}
u(t) = \kappa_e^m(x(t),\xi_t,\zeta_t) = u^*_{0|t}.
\end{align}
Similar to (\ref{eqn:etat}), the parameter $\zeta_t$ is determined using the previous optimal solution and the updated law is given below
\begin{align}\label{eqn:zeta}
\zeta_t = V^{\delta}(x(t-1),\pmb{u}^*_{t-1})- \beta J^{\delta}(x(t-1),\pmb{u}^*_{t-1}).
\end{align}
However, the parameter $\xi_t$ is updated using the optimal solution at time $t-m$ as shown below
\begin{align}\label{eqn:xi}
\xi_t = \max\{\tau \xi_{t-m},\zeta_{t-m+1}\} %V^{\delta}(x(t-m),\pmb{u}^*_{t-m})- \beta J^{\delta}(x(t-m),\pmb{u}^*_{t-m}) = \zeta_{t-m+1}
\end{align}
where $\tau\in [0,1)$ is some fixed scalar. Such a update law is used to obtain the property in (\ref{eqn:Vtm}). For the initialization, large numbers are chosen for $\zeta_0$ and $\xi_t$ for all $t\in \Z_0^{m-1}$. One possible way is to determine the maximal value function $V^{\delta}(x,\pmb{u})$ within the feasible domain. Let
\begin{subequations}
\begin{align}
V_{\max} =  \max\limits_{x,\pmb{u}} & V(x,\pmb{u})\\
\textrm{s.t.   } &x \in X, \pmb{u} \in \U(x).
\end{align}
\end{subequations}
At the initialization, let $\zeta_0=V_{\max}$ and $\xi_t = V_{\max}$ for all $t\in \Z_0^{m-1}$. Hence, we know that, for $t\in \Z_0^{m-1}$, $\mathbb{P}^{m}_{e}(x(t),\xi_t,\zeta_t)$ will reduce to the following problem
\begin{subequations}\label{eqn:EMPCLym1m}
\begin{align}
\bar{\mathbb{P}}^{m}_{e}(x(t),\zeta_t):\quad \min\limits_{\pmb{u}} &V_e(x(t),\pmb{u}):= \sum\limits_{k=0}^{N-1}l_e(x_k,u_k)  \\
\textrm{s.t.   } &\pmb{u} \in \U(x(t)),\\
& V^{\delta}(x(t),\pmb{u})- \beta J^{\delta}(x(t),\pmb{u}) \le \zeta_t. \label{eqn:V0l0Vtminus1}
\end{align}
\end{subequations}
At the initial state $x(0)$, $\mathbb{P}^{m}_{e}(x(t),\xi_t,\zeta_t)$ will further reduce to $\mathbb{P}_{e}(x(0))$. The overall scheme is summarized below.

\begin{algorithm}[H]
\caption{The economic MPC with a non-monotonic Lyapunov function}
\begin{algorithmic}[1]
\State \textit{Initialization}:  Set $t=0$, measure the initial state $x(0)$, let $\xi_0 = \zeta_0  =V_{\max} $, solve $ \mathbb{P}^{m}_{e}(x(t),\xi_t,\zeta_t)$ and obtain the optimal solution $\pmb{u}^*_0$. Apply the control law (\ref{eqn:kappam}) to system (\ref{eqn:xplus1AB}). Let $t:=t+1$ and go to Step 2.
\State Measure the current state $x(t)$, obtain $\zeta_t$ from (\ref{eqn:zeta}), and obtain $\xi_t$ from (\ref{eqn:xi}) when $t\ge m$ and set $\xi_t=V_{\max}$ when $t<m$.
\State Solve $\mathbb{P}^{m}_{e}(x(t),\xi_t,\zeta_t)$ and obtain the optimal solution $\pmb{u}^*_t$. Apply the control law (\ref{eqn:kappam}) to system (\ref{eqn:xplus1AB}).
\State Wait for next sampling time, let $t:=t+1$ and  go to Step 2.
\end{algorithmic}
\label{algo:EMPCm}
\end{algorithm}

\section{Recursive feasibility and stability}\label{sec:stability}
This section discusses the recursive feasibility and stability of the proposed economic MPC with $m>1$. As the constraint (\ref{eqn:Vx0uVtm}) is inactive for $t\in \Z_0^{m-1}$, the recursive feasibility is discussed in different cases: 1) $t=0$; 2) $1 \le t\le m-1$; 3) $t \ge m$. The following lemma shows that the feasibility of the initial state implies the feasibility of the next state.
\begin{lemma}\label{lem:P0}
Suppose $\mathbb{P}_{e}(x(0))$ has a feasible solution at time $t=0$. Let $\zeta_{1}$ be updated according to (\ref{eqn:zeta}) and the MPC control law (\ref{eqn:kappam}) is applied to the system (\ref{eqn:xplus1AB}). Then, $\bar{\mathbb{P}}^{m}_{e}(x(1),\zeta_{1})$ also has a feasible solution.
\end{lemma}
\textbf{Proof of Lemma \ref{lem:P0}:} Let $\pmb{u}_0^*$ denote the optimal solution of $\mathbb{P}_{e}(x(0))$ and $\pmb{x}_0^*:=\{x^*_{0|0},x^*_{1|0},\cdots,x^*_{N|0}\}$ be the associated predicted state sequence. Consider the shifted control sequence $\tilde{\pmb{u}}_{1}=:\{u^*_{1|0},\cdots,u^*_{N-1|0},\kappa_f(x^*_{N|0})\}$, from the arguments in Lemma \ref{lem:convergence}, we can see that the feasibility of $\mathbb{P}_{e}(x(0))$ implies that the shifted control sequence satisfies $\tilde{\pmb{u}}_{1}\in \U(x(1))$ and
\begin{align*}
&V^{\delta}(x(1),\tilde{\pmb{u}}_{1})- \beta J^{\delta}(x(1),\tilde{\pmb{u}}_{1})\\
\le &V^{\delta}(x(1),\tilde{\pmb{u}}_{1})  \le V^{\delta}(x(0),\pmb{u}^*_{0})- \beta J^{\delta}(x(0),\pmb{u}^*_{0}) = \zeta_{1}
\end{align*}
This guarantees the feasibility of $\bar{\mathbb{P}}^{m}_{e}(x(1),\zeta_{1})$. $\Box$

The recursive feasibility of (\ref{eqn:EMPCLym1m}) is stated below for $t\in \Z^{m-1}$.
\begin{lemma}\label{lem:feasibilitymbar}
Suppose $\bar{\mathbb{P}}^{m}_{e}(x(t),\zeta_t)$ has a feasible solution at time $t$ with $t\in \Z^{m-1}$. Let $\zeta_{t+1}$ be updated according to (\ref{eqn:zeta}) and the MPC control law (\ref{eqn:kappam}) is applied to the system (\ref{eqn:xplus1AB}). Then, $\bar{\mathbb{P}}^{m}_{e}(x(t+1),\zeta_{t+1})$ also has a feasible solution.
\end{lemma}
\textbf{Proof of Lemma \ref{lem:feasibilitymbar}:} The proof follows the same arguments in Lemma \ref{lem:P0}. Consider the optimal solution $\pmb{u}_t^*$ and the associated predicted state sequence $\pmb{x}_t^*$ at time $t$, the shifted control sequence $\tilde{\pmb{u}}_{t+1}=:\{u^*_{1|t},\cdots,u^*_{N-1|t},\kappa_f(x^*_{N|t})\}$ satisfies $\tilde{\pmb{u}}_{t+1}\in \U(x(t+1))$ and
\begin{align*}
&V^{\delta}(x(t+1),\tilde{\pmb{u}}_{t+1})- \beta J^{\delta}(x(t+1),\tilde{\pmb{u}}_{t+1})\\
\le &V^{\delta}(x(t+1),\tilde{\pmb{u}}_{t+1})  \le V^{\delta}(x(t),\pmb{u}^*_{t})- \beta J^{\delta}(x(t),\pmb{u}^*_{t}) = \zeta_{t+1}.
\end{align*}
This guarantees the feasibility of $\bar{\mathbb{P}}^{m}_{e}(x(t+1),\zeta_{t+1})$. $\Box$

The following lemma shows that the feasibility of (\ref{eqn:EMPCLym1m}) for $t\in \Z^{m-1}_0$ implies the feasibility of (\ref{eqn:EMPCLym}) for all $t\in \Z_m^{2m-1}$.

\begin{lemma}\label{lem:feasibilitymbarm}
Suppose $\bar{\mathbb{P}}^{m}_{e}(x(t),\zeta_t)$ has a feasible solution for all $t\in \Z^{m-1}_0$ with $\zeta_t$ being updated according to (\ref{eqn:zeta}). Let $\xi_t$ be updated according to (\ref{eqn:xi}). The MPC control law (\ref{eqn:kappam}) is applied to the system (\ref{eqn:xplus1AB}). Then, $\mathbb{P}^{m}_{e}(x(t),\xi_t,\zeta_t)$ also has a feasible solution for all $t\in \Z_m^{2m-1}$.
\end{lemma}
\textbf{Proof of Lemma \ref{lem:feasibilitymbarm}:} The feasibility of the constraint (\ref{eqn:V0l0Vtminus12}) follows the same arguments in Lemma \ref{lem:feasibilitymbar}. From this constraint, we can also know that $\zeta_{m} \le \cdots \le \zeta_{1}$. Consider the optimal solution $\pmb{u}_{m-1}^*$ at $m-1$, following the arguments in Lemma \ref{lem:convergence}, we can see that there exits a feasible shifted control sequence $\tilde{\pmb{u}}_{m}$ at time $m$ such that $\tilde{\pmb{u}}_{m}\in \U(x(m))$ and
\begin{align*}
V^{\delta}(x(m),\tilde{\pmb{u}}_{m})  \le& V^{\delta}(x(m-1),\pmb{u}^*_{m-1})  \\
&- \beta J^{\delta}(x(m-1),\pmb{u}^*_{m-1}) \\
=& \zeta_m,
\end{align*}
This, together with the fact that $\xi_m = \max\{\tau \xi_{0},\zeta_{1}\} \ge \zeta_1 \ge \zeta_m$, implies
\begin{align}
V^{\delta}(x(m),\tilde{\pmb{u}}_{m}) \le \xi_m
\end{align}
Hence the feasibility of $\mathbb{P}^{m}_{e}(x(m),\xi_m,\zeta_m)$ can be guaranteed. By repeating this process, we can show the feasibility of $\mathbb{P}^{m}_{e}(x(t),\xi_t,\zeta_t)$ for all $t\in \Z_m^{2m-1}$. $\Box$

Based on the result in Lemma \ref{lem:feasibilitymbarm}, the recursive feasibility of (\ref{eqn:EMPCLym}) is stated next.

\begin{lemma}\label{lem:feasibilitym}
Suppose $\mathbb{P}^{m}_{e}(x(\ell),\xi_{\ell},\zeta_{\ell})$ has a feasible solution for all $\ell \in \Z_{t-m+1}^{t}$ with $t\ge 2m-1$. Let $\xi_t$ and $\zeta_t$ be updated according to (\ref{eqn:xi}) and (\ref{eqn:zeta}) respectively for $t \ge m$. The MPC control law (\ref{eqn:kappam}) is applied to the system (\ref{eqn:xplus1AB}). Then, $\mathbb{P}^{m}_{e}(x(t+1),\xi_t,\zeta_t)$ also has a feasible solution.
\end{lemma}
\textbf{Proof of Lemma \ref{lem:feasibilitym}:} The proof just combines the arguments in the proofs of Lemma \ref{lem:feasibilitymbar} and \ref{lem:feasibilitymbarm} and hence is omitted. $\Box$

%follows similar arguments in Lemma \ref{lem:feasibilitymbarm}.  Consider that $V_t-J_t \le V_{t-1}-J_{t-1}\le \cdots \le V_{t-m+1}-J_{t-m+1}$ and the fact that there exits a feasible control sequence $\tilde{\pmb{u}}_{t+1}\in \U(x(t+1))$ such that
%\begin{align}
%V^{\delta}(x(t+1),\tilde{\pmb{u}}_{t+1})  \le V_{t}- J_{t},
%\end{align}
%we can obtain that
%\begin{align}
%V^{\delta}(x(t+1),\tilde{\pmb{u}}_{t+1}) \le V_{t-m+1}-J_{t-m+1}
%\end{align}
%Therefore, $\mathbb{P}^{m}_{e}(x(t+1))$ also has a feasible solution at time $t+1$.

%In order to discuss the stability, the following lemma is needed.
%
%\begin{lemma}\label{lem:ab}
%Let $\{a_k\}$ and $\{b_k\}$ be non-negative sequences. Suppose the following inequality is satisfied
%\begin{align}\label{eqn:abk}
%a_k \le \max\{\tau a_{k-1},a_{k-1} - b_{k-1}\}
%\end{align}
%with $\tau\in [0,1)$, then $\{a_k\}$ converges and $\lim\limits_{k\rightarrow \infty} \min\{a_{k},b_k\} = 0$.
%\end{lemma}
%\textbf{Proof of Lemma \ref{lem:ab}:} From (\ref{eqn:abk}), we know that
%\begin{align}
%a_k - a_{k-1} &\le \max\{\tau a_{k-1},a_{k-1} - b_{k-1}\} - a_{k-1} \nonumber\\
%&= \max\{\tau a_{k-1}-a_{k-1},a_{k-1} - b_k - a_{k-1}\} \nonumber\\
%&= \max\{(\tau-1) a_{k-1},- b_k \} \nonumber\\
%&= -\min\{(1-\tau) a_{k-1},b_{k-1} \}
%\end{align}
%This inequality implies that $\{a_k\}$ a monotonic non-increasing sequence bounded from below by 0. Hence, $\lim\limits_{k\rightarrow \infty}a_k$ exists and $\lim\limits_{k\rightarrow \infty} \min\{a_{k},b_k\} = 0$. $\Box$

With the results above, the recursive feasibility and stability is stated in the following theorem.

\begin{theorem}\label{thm:as}
Suppose $\mathbb{P}_{e}(x(0))$ has a feasible solution. Let $\xi_t$ and $\zeta_t$ be updated according to (\ref{eqn:xi}) and (\ref{eqn:zeta}) respectively with the initialization $\zeta_0 = V_{\max}$ and $\xi_t = V_{\max}$ for all $t\in \Z_0^{m-1}$. The MPC control law (\ref{eqn:kappam}) is applied to the system (\ref{eqn:xplus1AB}). Then, the following results hold with $m>1$.\\
(i) $\mathbb{P}^{m}_{e}(x(t+1),\xi_t,\zeta_t)$ is feasible for all $t\ge 0$.\\
(ii) $\lim\limits_{t\rightarrow \infty}V^{\delta}(x(t),\pmb{u}^*_{t}) = 0$ and $\lim\limits_{t\rightarrow \infty}J^{\delta}(x(t),\pmb{u}^*_{t}) = 0$.\\
(iii) The closed-loop system (\ref{eqn:xplus1AB}) controlled by the MPC control law (\ref{eqn:kappam}) asymptotically converges to $x_s$.
\end{theorem}
\textbf{Proof of Theorem \ref{thm:as}:} (i) This result follows from Lemma \ref{lem:P0}, \ref{lem:feasibilitymbar}, \ref{lem:feasibilitymbarm} and \ref{lem:feasibilitym}, as $\mathbb{P}_{e}(x(0))$ is feasible $\Rightarrow$ $\bar{\mathbb{P}}^{m}_{e}(x(t),\zeta_t)$ is feasible for $t\in \Z^{m-1}$ $\Rightarrow$ $\mathbb{P}^{m}_{e}(x(t),\xi_{t},\zeta_{t})$is feasible for $t \ge m$. \\
(ii) From (\ref{eqn:Vx0uVtm}), (\ref{eqn:zeta}) and (\ref{eqn:xi}), we know that, for all $t\ge 0$
\begin{align}
&V^{\delta}(x(t+m),\pmb{u}^*_{t+m})\nonumber\\
\le & \xi_{t+m} = \max\{\tau \xi_{t}, V^{\delta}(x(t),\pmb{u}^*_{t})- \beta J^{\delta}(x(t),\pmb{u}^*_{t}) \} \nonumber\\
\le & \max\{\tau \xi_{t}, \xi_{t}- \beta J^{\delta}(x(t),\pmb{u}^*_{t}) \} \nonumber\\
= & \max\{\xi_{t} - (1-\tau) \xi_{t}, \xi_{t}- \beta J^{\delta}(x(t),\pmb{u}^*_{t}) \} \nonumber\\
= & \xi_{t} - \min\{(1-\tau) \xi_{t},\beta J^{\delta}(x(t),\pmb{u}^*_{t})\}.\label{eqn:Vxi}
\end{align}
Consider that $\xi_{t} \ge V^{\delta}(x(t),\pmb{u}^*_{t})\ge J^{\delta}(x(t),\pmb{u}^*_{t})$, where the second inequality follows from (\ref{eqn:J}), we can know from (\ref{eqn:Vxi}) that
\begin{align}
\xi_{t+m} &\le \xi_{t} - \min\{(1-\tau) J^{\delta}(x(t),\pmb{u}^*_{t}),\beta J^{\delta}(x(t),\pmb{u}^*_{t})\} \nonumber\\
& = \xi_{t} - \min\{1-\tau,\beta\}J^{\delta}(x(t),\pmb{u}^*_{t}). \label{eqn:xitm}
\end{align}
Hence, $\{\xi_{t}\}$ satisfies the condition in (\ref{eqn:Vtm}). Let $\bar{\xi}^{\ell}_{k} = \xi_{km+\ell}$ and $\bar{J}^{\ell}_k = J^{\delta}(x(km+\ell),\pmb{u}^*_{km+\ell})$ for $k\ge0$ and $\ell\in \Z^{m-1}_0$. The inequality (\ref{eqn:xitm}) implies that
\begin{align}
\bar{\xi}^{\ell}_{k} \le \bar{\xi}^{\ell}_{k-1} - \min\{1-\tau,\beta\}\bar{J}^{\ell}_{k-1}.
\end{align}
This inequality implies that $\{\bar{\xi}^{\ell}_{k}\}$ a monotonic non-increasing sequence bounded from below by 0. Hence, $\lim\limits_{k\rightarrow \infty} \bar{\xi}^{\ell}_{k}$ exists and $\lim\limits_{k\rightarrow \infty} \bar{J}^{\ell}_{k} = 0$ for all $\ell\in \Z^{m-1}_0$. This means that $\lim\limits_{t\rightarrow \infty} J^{\delta}(x(t),\pmb{u}^*_{t}) = 0$. Finally, $\lim\limits_{t\rightarrow \infty} V^{\delta}(x(t),\pmb{u}^*_{t}) = 0$. \\
(iii) The asymptotic convergence to $x_s$ follows from (ii). $\Box$

%\subsection{Exponential stability}
%From theorem \ref{thm:as}, asymptotical convergence of the closed-loop system is guaranteed. To achieve stronger stability results, additional conditions are needed. If there exists a $\gamma>0$ such that $V_t \le \gamma \|x(t)\|^2$， it is possible to achieve exponential stability.
%
%\begin{lemma}
%For any $x\in X_f$, there exists a $\gamma>0$ such that $V^o(x)\le \gamma \|x\|^2$.
%\end{lemma}
%
%\begin{lemma}
%  For any $x\in X_f$, there exists a feasible $\pmb{u}$ to $\mathbb{P}^{m}_{e}(x)$ with $V(x_0,\pmb{u}) \le \gamma \|x\|^2$.
%\end{lemma}
%
%\begin{theorem}
%  Suppose the additional constraint $V(x_0,\pmb{u}) \le \gamma \|x\|^2$ is imposed on $\mathbb{P}^{m}_{e}(x)$ for any $x\in X_f$. Then, the closed-loop system (\ref{eqn:xplus1AB}) with the MPC control law (\ref{eqn:ukappa}) is exponentially stable.
%\end{theorem}

\section{Average economic performance}\label{sec:av}
The average performance of the proposed economic MPC scheme is analyzed in this section. Using the results above, the average asymptotic performance is stated below. This is similar to the economic MPC approaches in the literature\cite{ART:EDC14}.

\begin{theorem}\label{thm:performance}
Suppose $\mathbb{P}_{e}(x(0))$ has a feasible solution. Let $\xi_t$ and $\zeta_t$ be updated according to (\ref{eqn:xi}) and (\ref{eqn:zeta}) respectively with the initialization $\zeta_0 = V_{\max}$ and $\xi_t = V_{\max}$ for all $t\in \Z_0^{m-1}$. Then, the closed-loop system (\ref{eqn:xplus1AB}) with the MPC control law (\ref{eqn:kappam}) has an average cost that is no higher than that of the admissible optimal economic steady state.
\end{theorem}
\textbf{Proof of Theorem \ref{thm:performance}:} Let $\pmb{u}^*_t$ be the optimal control sequence obtained from Algorithm \ref{algo:EMPCm} at time $t\ge 0$ with the associated state sequence $\pmb{x}^*_t$. From the recursive feasibility(Property (i) of Theorem \ref{thm:as}), we know that the following control sequence is a feasible solution at time $t+1$
\begin{align}
\tilde{\pmb{u}}_{t+1}:=\{u^*_{1|t},\cdots,u^*_{N-1|t},\kappa_f(x^*_{N|t})\}.
\end{align}
The state sequence associated with $\tilde{\pmb{u}}_{t+1}$ is denoted by
\begin{align}
\tilde{\pmb{x}}_{t+1}:=\{x^*_{1|t},\cdots,x^*_{N|t},f(x^*_{N|t},\kappa_f(x^*_{N|t}))\}.
\end{align}
Using this feasible solution, we can obtain that
\begin{align*}
&V_e(x(t+1),\tilde{\pmb{u}}_{t+1}) - V_e(x(t),\pmb{u}^*_{t}) \\
= & l_e(x^*_{N|t},\kappa_f(x^*_{N|t}))-l_e(x(t),u(t)).
\end{align*}
Suppose $\pmb{u}^*_{t+1}$ is the optimal solution at time $t+1$, from the optimality, $V_e(x(t+1),\pmb{u}^*_{t+1})\le V_e(x(t+1),\tilde{\pmb{u}}_{t+1})$. Hence,
\begin{align}
&V_e(x(t+1),\pmb{u}^*_{t+1}) - V_e(x(t),\pmb{u}_{t}) \nonumber\\
\le & l_e(x^*_{N|t},\kappa_f(x^*_{N|t}))-l_e(x(t),u(t)).\label{eqn:Vextplus1}
\end{align}
From Property (ii) of Theorem \ref{thm:as}, we know that $\lim\limits_{t\rightarrow \infty}x^*_{N|t} = x_s$. Consider the continuity of the function $l_e(x,u)$, it can be obtained that $\lim\limits_{t\rightarrow \infty} l_e(x^*_{N|t},\kappa_f(x^*_{N|t})) = l_e(x_s,u_s)$ and $\lim\limits_{T\rightarrow \infty}\frac{1}{T+1} \sum\limits_{t=0}^{T} l_e(x^*_{N|t},\kappa_f(x^*_{N|t}))= l_e(x_s,u_s)$. Taking averages in both sides of (\ref{eqn:Vextplus1}) yields
\begin{align}
&\liminf\limits_{T\rightarrow +\infty} \frac{\sum\limits_{t=0}^{T}V_e(x(t+1),\pmb{u}^*_{t+1}) - V_e(x(t),\pmb{u}^*_{t})}{T+1} \nonumber\\
\le &\liminf\limits_{T\rightarrow +\infty}\frac{\sum\limits_{t=0}^{T}l_e(x^*_{N|t},\kappa_f(x^*_{N|t}))-l_e(x(t),u(t))}{T+1} \nonumber\\
= &l_e(x_s,u_s) - \limsup\limits_{T\rightarrow +\infty}\frac{\sum\limits_{t=0}^{T}l_e(x(t),u(t))}{T+1}. \label{eqn:liminf}
\end{align}
Consider that $X$ and $U$ are bounded, $V_e(x(t),\pmb{u}^*_{t})$ are also bounded for all $t\ge 0$. Hence,the left side of (\ref{eqn:liminf}) becomes
\begin{align*}
&\liminf\limits_{T\rightarrow +\infty} \frac{\sum\limits_{t=0}^{T}V_e(x(t+1),\pmb{u}^*_{t+1}) - V_e(x(t),\pmb{u}^*_{t})}{T+1} \\
= &\liminf\limits_{T\rightarrow +\infty} \frac{V_e(x(T+1),\pmb{u}^*_{T+1}) - V_e(x(0),\pmb{u}^*_{0})}{T+1} = 0.
\end{align*}
This, together with (\ref{eqn:liminf}), implies that
\begin{align}
\limsup\limits_{T\rightarrow +\infty}\frac{\sum\limits_{t=0}^{T}l_e(x(t),u(t))}{T+1} \le l_e(x_s,u_s).
\end{align}
$\Box$

\section{The application to HVAC systems}\label{sec:num}
\begin{figure}[H]
\centering
\includegraphics[width=0.9\linewidth]{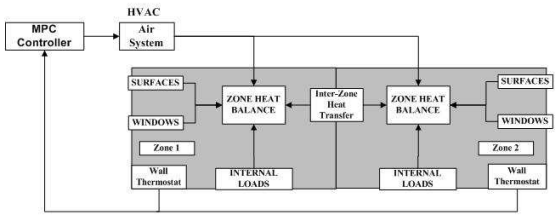}
\caption{Double-Zone Building Thermal Model}
\label{fig:model}
\end{figure}

This section discusses the application of the proposed economic MPC scheme to HVAC systems. We consider a virtual building composed of two adjacent zones as shown in Figure \ref{fig:model}. There is an air handling unit serving the two zones. The cool air is distributed by a fan and the flow rate of air supplied to each zone is controlled by the variable air volume boxes.

\subsection{Thermal model for the zones}
In this model, the temperature in each zone is assumed to be uniform. The thermal dynamics of each zone is described by a Resistive-Capacitive (RC) system \cite{ART:KMB13,INP:LMBJ12,INP:LLXJ16}. According to the zone heat balance equation, the thermal models of the two zones are given by
\begin{align}
c_i \dot{T}_i =& \frac{T_{j}-T_i}{R_{ij}} + \frac{T_{o}-T_i}{R^o_{i}}  \nonumber\\
&+ u_ic_p(T^s_{i}-T_i) + q_i, j\not= i, i = 1,2
\end{align}
where $T_i$ is the temperature of zone $i$, $T_o$ is the temperature of outside air, $c_i$ is the thermal capacitance of the air in zone $i$, $R_{ij}$ denotes the thermal resistances between zone $i$ and zone $j$, $R^o_i$ denotes the thermal resistance between zone $i$ and the outside environment, $c_p$ is the specific heat capacity of air, $T^s_{i}$ is the temperature of the supply air delivered to zone $i$, $u_i$ is the flow rate into zone $i$ and $q_i$ is the thermal disturbance from internal loads like occupants and lighting. As the temperature of the supply air is usually constant over short intervals of time, it is assumed to be fixed and known. The outside air temperature here is $T_o=32^\circ$C. The system parameters are given in Table \ref{tab:para}.

\begin{table}[H]
\centering
\renewcommand{\arraystretch}{1}
\begin{tabular}{|c|c|c|}
\hline
Symbol & Value & Units \\ \hline
$c_1=c_2$ & $9.163\times 10^3$ & $kJ/K$ \\ \hline
$c_p$ & $1.012$ & $kJ/(kg\cdot K)$ \\ \hline
$R_{12}=R_{21}$ & $14$ & $kW/K$ \\ \hline
$R^o_1 = R^o_2$ & $50$ & $kW/K$ \\ \hline
$T^s_{1}=T^s_{2}$ & $15$ & \textdegree{}C \\ \hline
$T_{o}$ & $32$ & \textdegree{}C \\ \hline
$q_1 = q_2$ & $4$ & $kW$ \\ \hline
\end{tabular}
\caption{System parameters} \label{tab:para}
\end{table}

We discretize the continuous-time system by the zero-order-hold method with the sampling time $\Delta t= 10 \textrm{min}$. The discretized model is given by
%\begin{align}
%T_i(t+1) = T_i(t) + \frac{\Delta t}{R_{ij}c_i}(T_{j}(t)-T_i(t)) + \frac{\Delta t}{R^o_{i}c_i}(T_{o}(t)-T_i(t))  + \frac{\Delta t c_p}{c_i}u_i(t)(T_{si}(t)-T_i(t)), j\not= i, i = 1,2
%\end{align}
\begin{align*}
x(t+1) = Ax(t)+ g(x(t))u(t) + d
\end{align*}
where $x := (x_1,x_2) := (T_1,T_2)$, $u = (u_1,u_2)$, $A=[0.9940 ~ 0.0047;0.0047 ~ 0.9940]$, $g(x(t))=[0.0663(16-x_1(t)) ~ 0; 0 ~ 0.0663(16-x_2(t))]$, $d = (0.3038,0.3038)$. The input constraint is $U:=\{(u_1,u_2): u_1 \ge 0, u_2 \ge 0, u_1 + u_2 \le 3.2\}$. In this experiment, the temperature set-points of zone $1$ and zone $2$ are $24^\circ$C and $25^\circ$C respectively. This means the asymptotic state constraint set is a singleton $X_\infty:=\{(x_1,x_2):T_1 = 24,T_2 = 25\}$. Hence, $x_s = (24,25)$ and $u_s=(0.4646,0.4020)$.

\subsection{Simulation results}
\begin{figure}[H]
\centering
\includegraphics[width=0.8\linewidth]{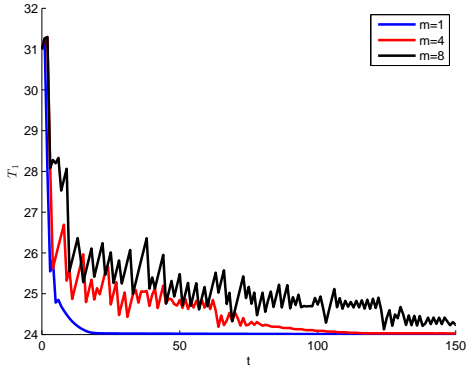}
\caption{The temperature of zone 1 for different $m$}
\label{fig:T1}
\end{figure}

\begin{figure}[H]
\centering
\includegraphics[width=0.8\linewidth]{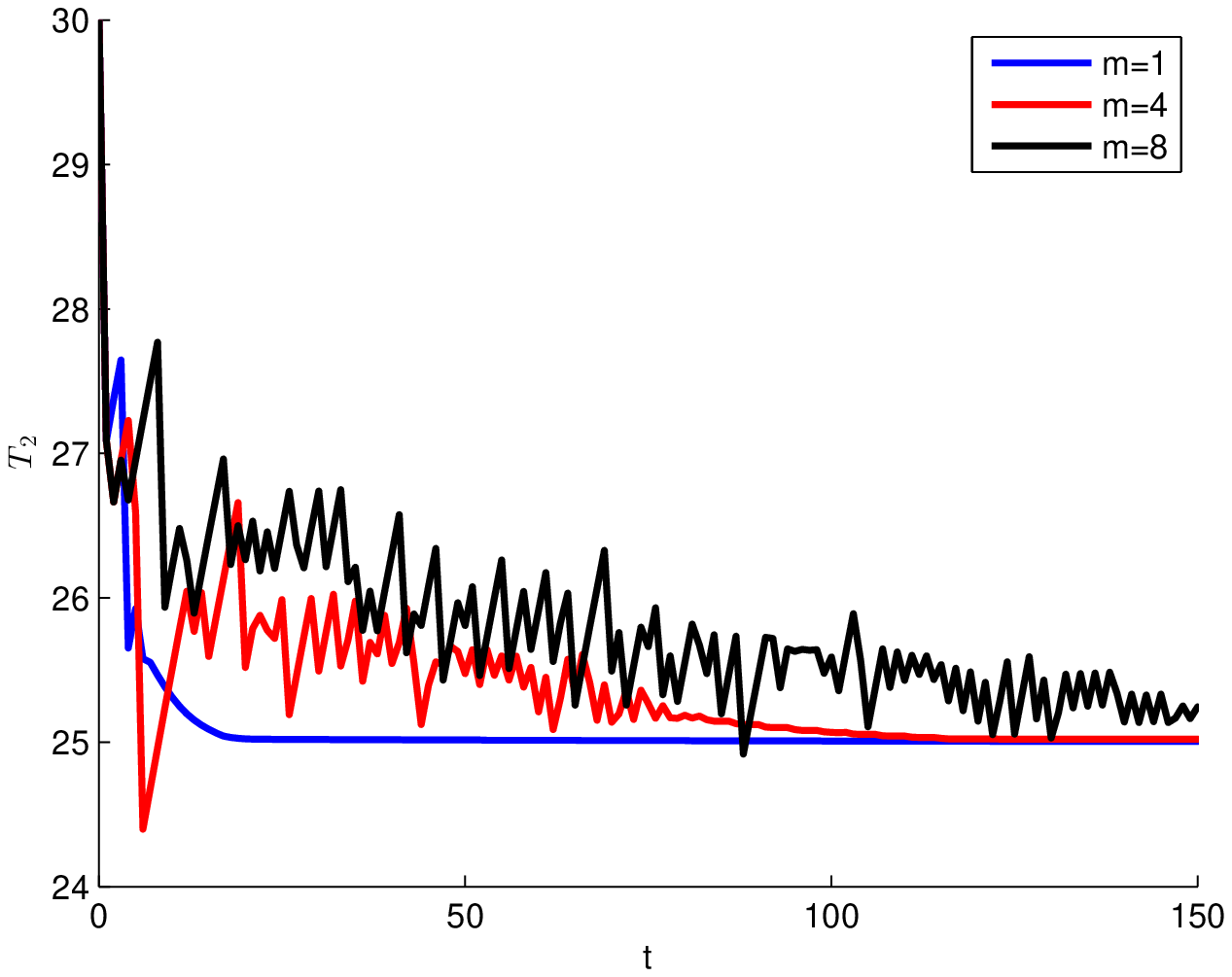}
\caption{The temperature of zone 2 for different $m$}
\label{fig:T2}
\end{figure}

\begin{figure}[H]
\centering
\includegraphics[width=0.8\linewidth]{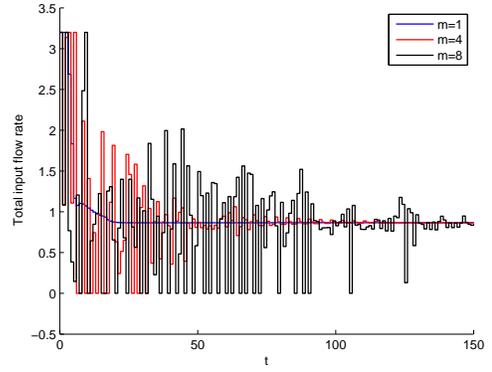}
\caption{Total input flow rates for different $m$}
\label{fig:u}
\end{figure}

\begin{figure}[H]
\centering
\includegraphics[width=0.8\linewidth]{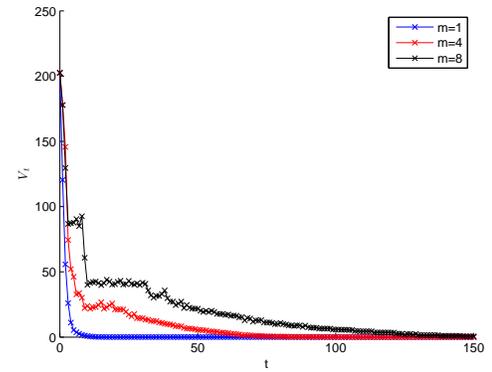}
\caption{The tracking value functions for different $m$}
\label{fig:vt}
\end{figure}

As shown in \cite{ART:KMB13,ART:MS12,INP:RMSR15}, the electrical power consumption can be approximated by 
\begin{align}
l_e(x,u) =& \bar{\kappa}(\sum\limits_{i} u_i)^3 + \frac{1}{\bar{\eta}_c}\sum\limits_{i} u_ic_p|T^s_{i}-x_i| \nonumber\\
&+ \frac{1}{\bar{\eta}_h}\sum\limits_{i} u_ic_p|T^h_i-x_i|
\end{align}
where $T^h_i$ is the heating coil set-point temperature, and the parameters $\bar{\kappa},\bar{\eta}_c$ and $\bar{\eta}_h$ capture the energy transfer efficiency. These parameters are chosen to be $T^h_1=T^h_2=32^\circ$C, $\bar{\eta}_c=4$ and $\bar{\eta}_h=0.9$. The tracking cost functions are in the form of
\begin{align}\label{eqn:TMPCcost}
\begin{aligned}
l(x,u) &= \|x-x_s\|_Q^2 + \|u-u_s\|_R^2\\
l_f(x) &= \|x-x_s\|_P^2
\end{aligned}
\end{align}
where $Q,R,P \succ 0 $. Let $Q = I$ and $R=I$. The terminal control law is $\kappa_f(x) = K(x-x_s)+u_s$ with $K = [0.6947 ~ 0.0059;0.0061 ~ 0.6818]$. The horizon length of MPC is set to be $N=5$. $P$ is determined such that (\ref{eqn:P}) is satisfied with $\delta(x,u) = 10^{-4}(\|x-x_s\|^2 + \|u-u_s\|^2)$ and $\gamma(x) = 10^{-4}\|x-x_s\|^2$. Set $\beta=1$ and $\tau=0.6$. Let the initial state be $x(0) = [31 ~ 30]^T$. The involved optimization problems are solved using the solver IPOPT \cite{ART:WB06} with the interface OPTI toolbox \cite{INP:JD12}. For different $m$, the temperatures of the two zones are shown in Figure \ref{fig:T1} $\&$ \ref{fig:T2} . As it is expected, a smaller $m$ provides faster convergence to the set-point. The total flow rate is $u_1+u_2$. Figure \ref{fig:u} shows the total flow rates for different $m$. We can see that the total rate is small when $m$ is large. The convergence curves of the tracking value functions are also given in Figure \ref{fig:vt} with $V_t = V^{\delta}(x(t),\pmb{u}^*_t)$. From these curves, it can be seen that the tracking value function does not monotonically decrease for $m>1$. When $m=1$, it can be considered as the monotonically decreasing case in \cite{ART:HWS15,ART:Z15,ART:HSY16}.

%\begin{figure}[H]
%\centering
%\subfigure[]
%  {\includegraphics[width=0.8\linewidth]{T1.eps}} \vfil
%  \subfigure[]
%  {\includegraphics[width=0.8\linewidth]{T2.eps}} \\
%  \caption{Temperatures of the two zones for different $m$}
%\label{fig:T}
%\end{figure}

The total energy consumption for $24$ hours is shown in the following table. This table also provides the comparison with standard tracking MPC with the objective being in the form of (\ref{eqn:TMPCcost}). For the case of $m=1$, the energy consumption is reduced by $1.4\%$ compared to tracking MPC. As $m$ increases, more energy can be saved. For $m=8$, there is a significant reduction in energy consumption, which is more than $20\%$.

\begin{table}[H]
\centering
\renewcommand{\arraystretch}{1}
\begin{tabular}{|c|c|c|c|c|}
\hline
MPC scheme & $m=1$ & $m=4$ & $m=8$ & Tracking MPC \\ \hline
Consumption($kWh$) & $240.3$ & $219.1$ & $194.2$ & $243.7$ \\ \hline
\end{tabular}
\caption{The total energy consumption for $24$ hours} \label{tab:cost}
\end{table}

\subsection{Average economic cost}
\begin{figure}[H]
\centering
\includegraphics[width=0.8\linewidth]{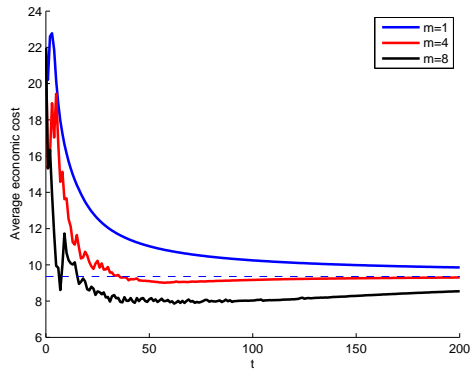}
\caption{The average economic cost functions for different $m$: the dashed line denotes the value of $l_e(x_s,u_s)$}
\label{fig:cost}
\end{figure}

The rest of this section discusses the economic performance of the proposed economic MPC approach. The average economic performance at time $t$ is measured by the following average economic cost function
\begin{align*}
\frac{\sum\limits_{k=0}^{t}l_e(x(k),u(k))}{t+1}
\end{align*}
where $x(k)$ and $u(k)$ are the true state and control of the close-loop system. Figure \ref{fig:cost} shows the average economic cost functions for different $m$. This figure verifies the statement in Theorem \ref{thm:performance} that the average performance is no worse than that of the admissible optimal economic steady state. In addition, we can also see the average economic cost is low when $m$ is large, which is consistent with the results in Figure \ref{fig:u}.

\section{Conclusion}\label{sec:con}
An economic MPC scheme is proposed by the use of Lyapunov-based constraints. Like the existing approaches, these constraints will enforce a Lyapunov decrease and ensure the asymptotical convergence to the steady state. Unlike them, relaxed Lyapunov-based constraints are considered to enhance the economic performance. Because of such constraints, the Lyapunov function decreases after a fixed number of steps. The decrease speed can be controlled by tuning this fixed number of steps. A virtual building example composed of two zones is presented to demonstrate the performance of the proposed economic MPC scheme. The trade-off between convergence speed and economic performance can be observed in the numerical results.

\bibliographystyle{unsrt}
{\footnotesize
\bibliography{Reference}}

%% not want to use model2-names.bst.



%% \bibitem must have one of the following forms:
%%   \bibitem[Jones et al.(1990)]{key}...
%%   \bibitem[Jones et al.(1990)Jones, Baker, and Williams]{key}...
%%   \bibitem[Jones et al., 1990]{key}...
%%   \bibitem[\protect\citeauthoryear{Jones, Baker, and Williams}{Jones
%%       et al.}{1990}]{key}...
%%   \bibitem[\protect\citeauthoryear{Jones et al.}{1990}]{key}...
%%   \bibitem[\protect\astroncite{Jones et al.}{1990}]{key}...
%%   \bibitem[\protect\citename{Jones et al., }1990]{key}...
%%   \harvarditem[Jones et al.]{Jones, Baker, and Williams}{1990}{key}...
%%


% \end{thebibliography}

\end{document}